\RequirePackage{ifpdf}
\documentclass{JHEP3}
\pdfoutput=1
\usepackage{slashed}
\usepackage{graphicx}
\usepackage{comment}

\newcommand{\lambdab}{{\overline{\lambda}}}
\newcommand{\chib}{{\overline{\chi}}}

\newcommand{\psib}{{\overline{\psi}}}

\newcommand{\xib}{{\overline{\xi}}}
\newcommand{\Tr}{{\rm Tr\;}}

\def\bec{\begin{center}}
\def\eec{\end{center}}
\def\beq{\begin{equation}}
\def\eeq{\end{equation}}
\def\bea{\begin{eqnarray}}
\def\eea{\end{eqnarray}}

\title{A Complete Lattice Technicolor Model}

\author{Simon Catterall and Aarti Veernala,\\
Department of Physics, Syracuse University, Syracuse, NY 13244, USA}

\abstract{We construct a lattice gauge theory using reduced staggered fermions  and gauge fields which
provides a non-perturbative realization of a {\it complete} technicolor model; one which treats both strong and weakly coupled gauge
sectors on an equal footing.
We show that the model is capable of developing a Higgs phase at  non zero lattice spacing via the formation of
fermion condensates. We further show that while the broken symmetry associated with this phase has a vector character in the lattice theory
it is realized as an axial symmetry in the 
continuum limit in agreement with the Vafa Witten theorem. We discuss our result in the context of universality}


\begin{document}

\section{Introduction}

The idea that the Higgs mechanism can occur through the formation of fermionic condensates is an
attractive one when constructing many theories of Beyond Standard Model (BSM) physics and finds application in technicolor, composite Higgs models,
tumbling and grand unification schemes
\cite{TC-intro,TC-intro2, Raby:1979my, gut}. 
Lattice realizations of these scenarios thus potentially give a rigorous setting for understanding 
how non-perturbative dynamics in models without elementary scalars can
spontaneously break gauge symmetries and potentially can give us new tools to analyze such theories.

In a lattice theory Elitzur's theorem \cite{Elitzur:1975im} guarantees that any
condensate which is not invariant under the gauge symmetry must necessarily have vanishing
expectation value. Instead, the gauge invariant way to understand spontaneous gauge symmetry
breaking and the operation of the
Higgs mechanism in such theories is that it proceeds via the condensation of a {\it gauge invariant} four
fermion operator of the form  
$\phi^\dagger \phi$  where $\phi$ is a fermion bilinear
which carries a non-trivial representation of the gauge group.
One can think of this bilinear as a composite Higgs field.

In this paper we construct a lattice realization of these ideas using a set of two {\it massless} staggered lattice fermions.
Each of these fields can be {\it reduced} by simply restricting one field to even parity sites and the
other to sites of odd parity.
The key observation is that in the absence of single site mass terms
the kinetic terms for each of these reduced staggered fermions may 
be gauged independently. 
We will use this observation to
construct a class of lattice technicolor theories in which the gauge forces factorize into strong and weakly interacting sectors with
both reduced fields coupling identically to the strong interaction but differing in their weak interactions \cite{us}.
In the limit in which the weak gauge couplings are set to zero the strong interactions generate
the usual single site condensate characteristic of staggered fermions. This condensate, which couples the
two reduced fields, is a strong interaction
singlet but will spontaneously break a subgroup of the weak
symmetries.  Furthermore, we show that while the broken symmetries have
a vector character at non-zero lattice spacing, they are to be interpreted as axial
symmetries in the continuum limit. This result is then compatible with the Vafa Witten
theorem which forbids the spontaneous breaking of vector symmetries \cite{Vafa:1984xg}. This is a necessary (but not sufficient)
condition for this phase to survive the continuum limit.

Once the weak symmetries are gauged these bilinear condensates must remain zero in accord with
Elizur's theorem but instead, a non-trivial
four fermion condensate is formed. The appearance of this condensate signals that the system has entered a Higgs phase
in which the would be Goldstone bosons corresponding to
the broken weak symmetries disappear from the spectrum together with a discontinuous change
in the free energy of a weakly charged static source.

We first describe the details of the construction and then present results showing the spontaneous breaking of the weak symmetries
in the case where the weak gauge coupling is set to zero. We then describe the arguments that are needed
to show that the broken symmetries are to be interpreted as axial
symmetries in the continuum limit. Finally, we then switch on the weak gauge  coupling and show evidence that this 
breaking pattern persists and that a Higgs phase forms. We end with a discussion of possible continuum limits in the model.

\section{Lattice model}

We start with two staggered fields $\chi$ and $\xi$. After restricting them to odd/even sites, we can define new fields $\psi$ and $\lambda$ as :

\begin{eqnarray}
\psib_+(x)=\frac{1}{2}\left(1+\epsilon(x)\right)\chib(x), \; \; \; \; \; \; \; \; \lambdab_-(x)=\frac{1}{2}\left(1-\epsilon(x)\right)\xib(x)\nonumber\\
\lambda_+(x)=\frac{1}{2}\left(1+\epsilon(x)\right)\xi(x), \; \; \; \; \; \; \; \;  \psi_-(x)=\frac{1}{2}\left(1-\epsilon(x)\right)\chi(x)
\end{eqnarray}
where the parity of a lattice site is given by $\epsilon(x)=\left(-1\right)^{\sum_{\mu=1}^4 x_\mu}$.
The fields $\psi$ and $\lambda$ are termed  {\it reduced} staggered fermions since each contains half the
number of degrees of freedom of the usual staggered fermion and corresponds to
two rather than four Dirac fermions in the continuum limit \cite{Smit, Golterman}. The resultant lattice action can then be written as,

\beq
S=\sum_{x,\mu}\eta_\mu(x)\psib_+(x)\left(\psi_-(x+\mu)-\psi_-(x-\mu)\right)+
\sum_{x,\mu}\eta_\mu(x)\lambdab_-(x)\left(\lambda_+(x+\mu)-\lambda_+(x-\mu)\right)
\eeq
where the phase 
$\eta_\mu(x)$ is the usual staggered quark phase given by
\beq  \eta_{\mu}(x) = (-1)^{\sum_{i=1}^{\mu - 1} x_{i}} . \eeq
Since the fields $\lambda$ and $\psi$ in the action are uncoupled we can take them to transform
in different representations
of one or more internal symmetry groups.  In this paper we will assume a gauge
symmetry of the form $SU(N)\times SU(M)\times SU(M)$ with the reduced staggered fields
$\psi$ and $\lambda$ transforming in the $(\Box,\Box,1)$ and $(\Box,1,\Box)$ representations (a similar
construction was used in an earlier work by Banks et. al. \cite{Banks:1982gt}).
We will take
the $SU(N)$  gauge coupling to be large
while the two $SU(M)$'s are assumed weakly coupled  and act on the fields in the following way:
\begin{eqnarray}
\psib_+ \to \psib_+ G^\dagger, \; \;  \; \; \; \;\psi_- \to  G\psi_- ,  \; \; \; \; \; \; \lambdab_- \to \lambdab_-H^\dagger , \; \; \; \; \; \; \lambda_+&\to& H\lambda_+
\end{eqnarray}
where $G$ and $H$ denote the corresponding weak symmetry transformations and we have suppressed the variation of the fields under  the strong symmetries.
Notice that it is impossible to write down a single site mass term
that preserves all the weak symmetries.  The usual  staggered mass term
\beq
\psib_+(x)\lambda_+(x)+\lambdab_-(x)\psi_-(x)\eeq  is {\it not} invariant but instead
transforms as a bifundamental under the weak groups.
In the absence of such a mass term all the gauge symmetries can be made local by inserting appropriate gauge links
between the $\psi$ and $\lambda$ fields on neighboring sites.
The gauging of
the lattice kinetic term is given explicitly as
\begin{eqnarray}
S_K=&\sum_{x,\mu}&\psib_+(x)\left(U_\mu(x)V_\mu(x)\psi_-(x+\mu)-U^\dagger_\mu(x-\mu)V^\dagger_\mu(x-\mu)\psi_-(x-\mu)\right)\nonumber \\
&\sum_{x,\mu}&\lambdab_-(x)\left(W_\mu(x)V_\mu(x)\lambda_+(x+\mu)-W^\dagger_\mu(x-\mu)V^\dagger_\mu(x-\mu)\lambda_+(x-\mu)\right),
\label{staggered_kinetic}
\end{eqnarray} where $V_{\mu}(x)$ is the lattice gauge field for the
strong interactions and  $U_{\mu}(x)$ and $W_\mu(x)$ correspond to the weak gauge groups.

While there are no single site fermion bilinears that are gauge invariant it is nonetheless
possible to write down a gauge invariant lattice four fermion term:
\beq
\sum_{\mu} \phi^\dagger_{+}(x)U_{\mu}(x)W^*_\mu(x)\phi_{-}(x),
\eeq 
where the composite Higgs field $\phi(x)$ (a strong interaction singlet) is given by
\begin{eqnarray}
\phi^\dagger_+(x)=\psib_+(x)\lambda_+(x),  \; \; \; \; \; \; 
\phi_-(x)=\lambdab_-(x)\psi_-(x)
\end{eqnarray}
If a condensate of $\phi^\dagger \phi$ develops it will signal the appearance of a lattice phase
in which a subset of the weak interactions are Higgsed. This subset can be 
determined in the limit where the weak coupling is set to zero; in this case one expects
the
strong interaction to break the weak symmetries down to their diagonal
subgroup
\beq
SU_D(M)={\rm diag}\left( SU(M)\times SU(M)\right) \eeq
The Higgsed gauge fields will thus correspond to the broken generators
\beq
\tau_a=\tau^1_a-\tau^2_a\eeq
with $\tau^1$ and $\tau^2$ the generators of the two weak groups respectively.

In order to see such a Higgs phase, we need to add a small perturbation to the above action via an auxiliary field $\phi$. 
\beq
\delta S=g\sum_x \left(\phi(x)\psib_+(x)\lambda_+(x)+\phi^\dagger(x)\lambdab_-(x)\psi_-(x)\right)
\label{fourfermi}
\eeq
This field $\phi(x)$ is a local field which is needed to couple the two reduced fermions
together and must transform under
the gauge symmetry as a bifundamental in the weak groups in order that the perturbation is
gauge invariant.
To render the path integral well defined after
integration over $\phi(x)$ one must
then also add a suitable action for $\phi(x)$. We choose an additional simple term  $\sum_x \phi^\dagger(x)\phi(x)$.
The effect of these Yukawa  terms is to add a small gauge invariant  four fermion interaction to the action that
favors the conjectured symmetry breaking pattern. Spontaneous breaking of the 
global weak symmetries occurs iff the fermion bilinear takes a vacuum expectation
value in the thermodynamic limit as the auxiliary coupling is sent to zero.

Finally to build the entire model requires the addition of Wilson plaquette terms for
the three independent lattice gauge fields. 
In the rest of the paper will consider the case where the weak symmetry groups are $SU(2)$ while the strong symmetry
corresponds to $SU(3)$.

\section{Strong dynamics and chiral symmetry breaking}
\begin{figure}
\begin{center}
\includegraphics[angle=270, width=0.75\textwidth]{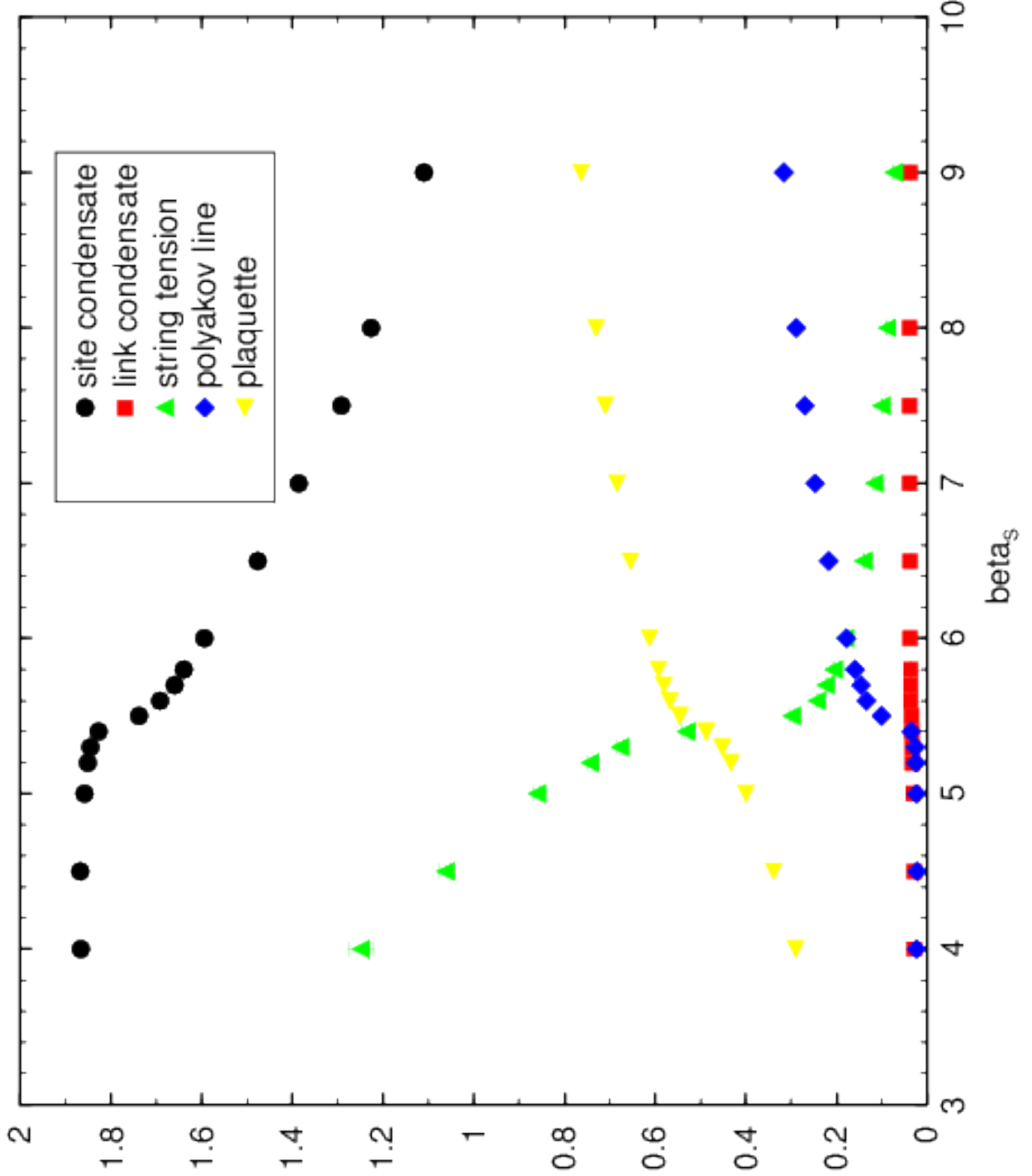}\end{center}
\caption{\label{globalbeta} Scan of observables with $\beta_S$ with $g = 0.1$ and $L=4$}
\end{figure} 

Let us first examine the case where the weak gauge coupling is set to zero and the weak symmetry is purely global.
Such a scenario can be realized by freezing the auxiliary scalar $\phi$ and the
gauge fields $U_\mu$ and $W_\mu$ to $\phi$ to the unit
matrix. The auxiliary Yukawa term described above
now acts as an explicit source term for the 
fermion bilinear 
\beq
\Sigma_S=\left(\psib_+(x)\lambda_-(x)+\lambdab_-(x)\psi_-(x)\right)
\eeq
We will be interested in monitoring the corresponding
single site condensate since a non-zero vacuum expectation value for this observable in the limit  $g\to 0$ would
indicate a spontaneous breaking of $SU(2)\times SU(2)$ down to its diagonal $SU(2)$ subgroup. However, this
is not the only fermion bilinear one can add to the action; it is also possible to construct a single link operator $\Sigma_L$ which
is given on even parity sites by the expression\footnote{Our action also contains the corresponding
odd site bilinear which is given by a similar
expression. We observe that the vacuum expectation of the odd site operator  matches its even site companion as one would
expect}
\beq
\Sigma_L=\frac{1}{8}\sum_\mu\left(\psib_+(x)V_\mu(x)\lambda(x+\mu)+V^\dagger(x-\mu)\lambda(x-\mu)\right)\eeq
A vacuum expectation value for this quantity {\it does not} break the weak symmetries. To examine which condensate is
preferred and hence whether the weak symmetries are broken we have also added an explicit source term for this single link bilinear to the lattice action with the {\it same}
coupling $g$ as the single site operator.

\begin{figure}
\begin{center}
\includegraphics[angle=270,width=0.75\textwidth]{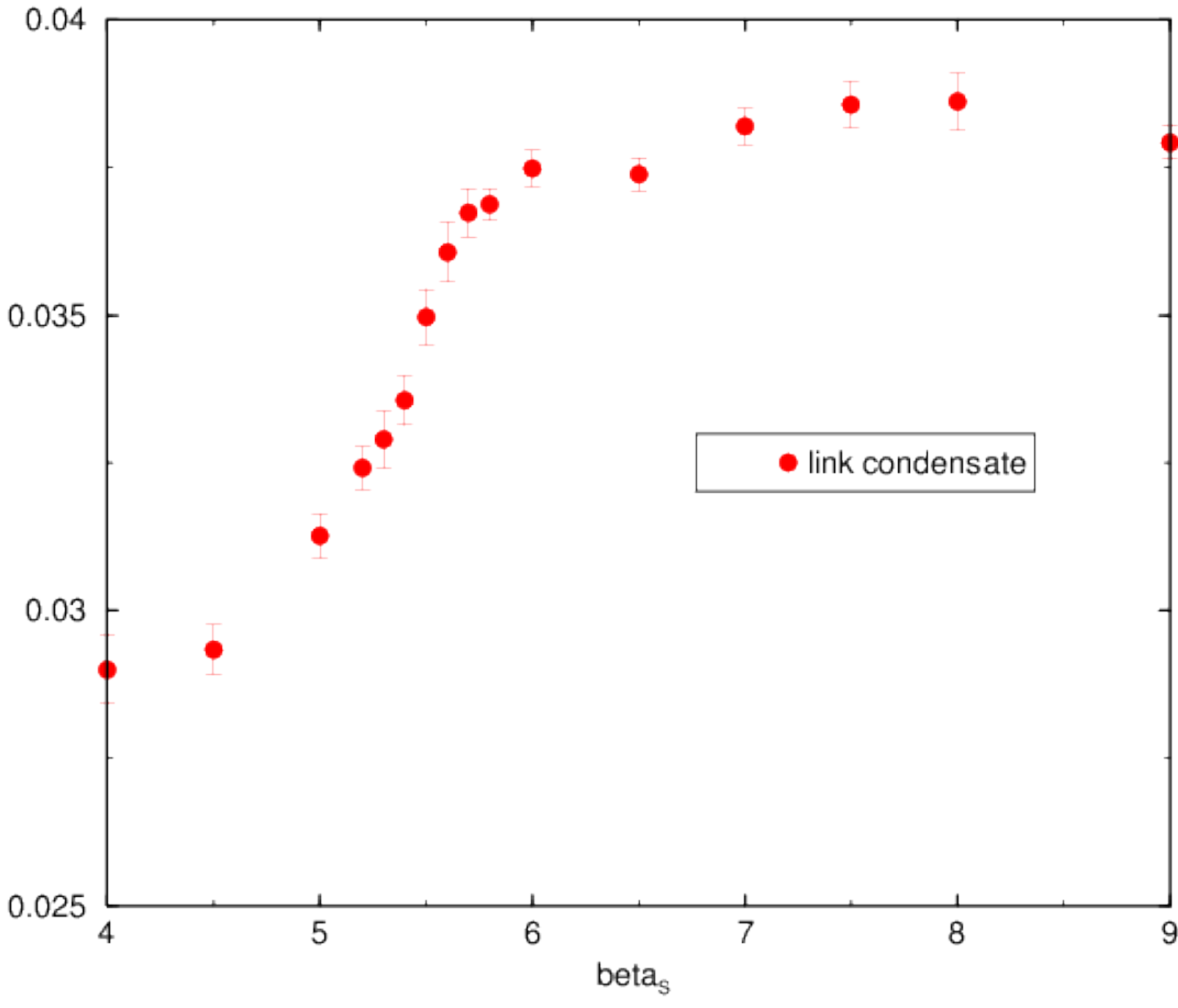}\end{center}
\caption{\label{single} $\Sigma_L$  vs $\beta_S$ with $g = 0.1$ and $L=4$}
\end{figure}

To reveal the phase structure of the model we first perform a scan in the strong coupling $\beta_S$. In fig.~\ref{globalbeta} we show
the average plaquette, Polyakov line, string tension, single site and single link condensates versus $\beta_S$ for fixed auxiliary
coupling $g=0.1$ (which can be  now interpreted as a mass parameter $m$). A single lattice volume $V=4^4$ is shown. All the data are consistent with a transition or crossover
region around $\beta_S=5.4$ which separates the confining and chirally broken phase expected of this $SU(3)$ gauge theory from a deconfined
regime encountered for sufficiently large $\beta_S$ on a finite lattice volume. Furthermore, the single site condensate  clearly dominates
over the single link condensate.
Indeed, in fig.~\ref{single} we show a magnified
plot of the single link condensate; not only is $\Sigma_S/\Sigma_L>50$ 
but $\Sigma_L$ is {\it suppressed} as the system enters the chirally broken regime in contrast to
$\Sigma_S$ which is enhanced. This should not be surprising; it is precisely the single site
condensate that is used as an order parameter for chiral symmetry breaking in simulations of lattice QCD using staggered fermions.

\begin{figure}
\begin{center}
\includegraphics[width=0.75\textwidth]{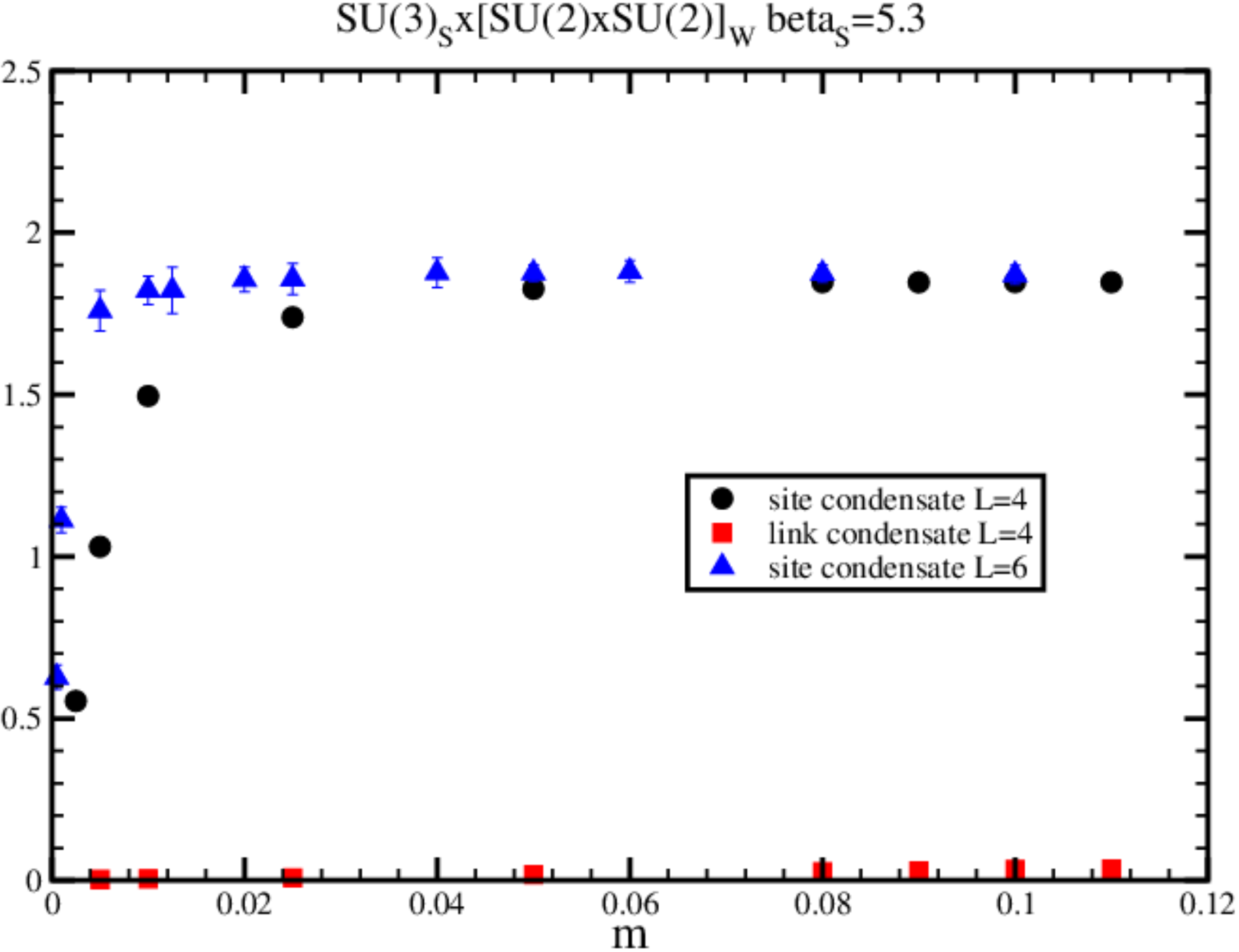}\end{center}
\caption{\label{gscan} Condensates vs $m$ for $L=4$ and $L=6$}
\end{figure}

To verify that this is truly a spontaneous breaking of $SU(2)\times SU(2)$ we have also performed a scan of the condensates at fixed $\beta_S=5.3$ varying
the auxiliary coupling $g\equiv m$. The data shown in fig.~\ref{gscan} shows that the single site condensate dominates at all $g$ with $\Sigma_S$ approaching a constant  as
$g$ is decreased before finally turning over to rapidly approach zero for $g=0$. Notice that this plateau region extends to smaller $g$ as the volume
increases consistent with a non-zero value of the condensate in the thermodynamic limit even for vanishing source  $g= 0$. The results thus
strongly support the idea that the global $SU(2)\times SU(2)$ symmetry is indeed spontaneously broken to its diagonal subgroup
in the large volume continuum limit for non zero lattice spacing.

At first sight this is puzzling; the weak symmetries have a vector-like character in the lattice theory while the Vafa-Witten theorem would appear to prohibit the
spontaneous breaking of such vector-like symmetries. In the next section we will show how to resolve this apparent contradiction.

\section{Axial symmetries in the continuum limit}
\label{global}
Since each reduced staggered field contributes 2 Dirac fermions in the continuum
limit  the theory will contain eight Dirac fields
with global symmetry $SU_V(8)\times SU_A(8)$.  As a result of strong interactions
we expect that this symmetry will break according to the pattern
\beq
SU_V(8)\times SU_A(8)\to SU_V(8). 
\label{su8}
\eeq
The $SU(2)\times SU(2)$ weak symmetries of the staggered lattice theory 
constitute a subgroup of this continuum global symmetry. The precise embedding of these
weak symmetries in the continuum limit can be obtained by the following argument.
We start by assuming that each staggered field yields 2 Dirac fermions in the
continuum eg. $\psi_1\to \{\Psi_1^1,\Psi_1^2\}$ with the upper indices reflecting this extra factor of two. We can then arrange these
continuum fields into an eight component vector
\beq\left(\begin{array}{cccccccc}
\Psi^1_1 & \Lambda^1_1 & \Psi^1_2 & \Lambda^1_2 &
\Psi^2_1 & \Lambda^2_1 & \Psi^2_2 & \Lambda^2_2 \end{array}\right)\eeq
In this representation the {\it broken}  generators $\tau_a =\tau^1_a-\tau^2_a,\,a=1\ldots 3$ take the explicit form
\beq
\begin{array}{ccc}
\tau_1=\left(\begin{array}{cc}
0&\sigma_3\\
\sigma_3&0\end{array}\right)\times I_2\qquad &
\tau_2=\left(\begin{array}{cc}
0&i\sigma_3\\
-i\sigma_3&0\end{array}\right)\times I_2 \qquad &
\tau_3=\left(\begin{array}{cc}
\sigma_3&0\\
0&-\sigma_3\end{array}\right)\times I_2
\end{array}
\eeq
where the two dimensional unit matrix $I_2$ represents this extra factor
of two degeneracy associated to the upper indices of $\Psi,\Lambda$. 
We will neglect the $I_2$ factor in what follows since it enters trivially in our analysis.
On the lattice the usual single site fermion condensate takes the form
\beq
\sum_{a=1}^2\psib^a_+\lambda^a_++\lambdab^a_-\psi^a_-
\label{cond}\eeq
corresponding to two independent staggered fermion condensates. This clearly
breaks the lattice $SU(2)\times SU(2)$ symmetry down to its diagonal subgroup. 
We then expect that this breaking pattern corresponds to a continuum condensate of the form
\beq
\Sigma=\left(\begin{array}{cc}
\sigma_1&0\\
0&\sigma_1\end{array}\right)\times I_2
\eeq
Standard universailty arguments tell us that it should be possible to change basis for our fermion fields to force this condensate to take
the canonical flavor symmetric form $\Sigma=I_8$. To accomplish this first diagonalize $M\to P^\dagger MP$ using the
unitary transformation
\beq
P=\left(\begin{array}{cc}
\frac{1}{\sqrt{2}}\left(\sigma_3+\sigma_1\right)& 0\\
0& \frac{1}{\sqrt{2}}\left(\sigma_3+\sigma_1\right)\end{array}\right)\eeq
This results in a condensate of the form
\beq
\Sigma^\prime=\left(\begin{array}{cc}\sigma_3&0\\
0&\sigma_3\end{array}\right)\eeq
To transform this to the unit matrix we employ the non-anomalous chiral transformation $M^\prime \to QM^\prime Q$
with
\beq
Q=\left(\begin{array}{cc}
D&0\\
0&D^\dagger\end{array}\right)\eeq
where the $2\times 2$ matrix
 \beq
D=\left(\begin{array}{cc}
1&0\\
0& i\gamma_5\end{array}\right)\eeq
To find the explicit form of the broken generators in this new basis we transform them according to
the rule $Q^\dagger P^\dagger \tau_a PQ$.  It is straightforward to show that
the broken generators acquire an {\it axial} character in the new basis (we have reinserted the $I_2$ at this point)
\beq
\begin{array}{ccc}
\tau^\prime_1=\gamma_5\left(\begin{array}{cc}
0&-i\sigma_1\\
i\sigma_1&0\end{array}\right)\times I_2\qquad &
\tau^\prime_2=\gamma_5\left(\begin{array}{cc}
0&\sigma_1\\
\sigma_1&0\end{array}\right)\times I_2\qquad &
\tau^\prime_3=\gamma_5\left(\begin{array}{cc}
\sigma_2&0\\
0&-\sigma_2\end{array}\right)\times I_2
\end{array}
\eeq

This confirms our earlier arguments that the broken generators do indeed correspond to axial symmetries in the continuum limit.
We see that the staggered fermion action we use picks out a particular breaking direction corresponding to a specific embedding of the weak symmetries
into the global symmetry group. The fact that the broken symmetries are axial in the continuum limit is
a necessary condition, according to
the Vafa-Witten theorem, for this broken phase of
the lattice theory to survive the continuum limit\footnote{The reduced staggered theory has a sign problem for
non zero lattice spacing. This allows a condensate that breaks vector symmetries to exist away from the continuum limit}.

One should contrast this
with the situation in the continuum where the canonical chiral symmetry breaking pattern given in eqn.~\ref{su8} 
gives rise to an infinite number of degenerate vacua related by
an $SU(8)$ transformation. Depending on the vacuum that is picked any given $SU(2)$ subgroup may be broken or left intact
by the corresponding vacuum condensate.

\section{Weakly gauging the $SU(2)\times SU(2)$ symmetries}

\begin{figure}
\begin{center}
\includegraphics[height=95mm]{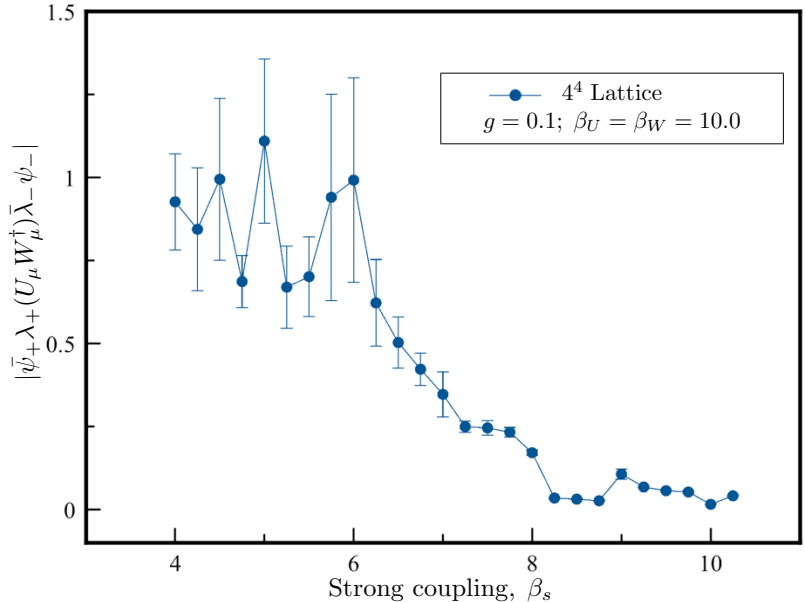}\end{center}
\caption{\label{nonsterile1} Absolute value of the four fermion condensate vs $\beta_{S}$ for auxiliary Yukawa coupling g = 0.1}
\end{figure}

\begin{figure}
\begin{center}
\includegraphics[height=95mm]{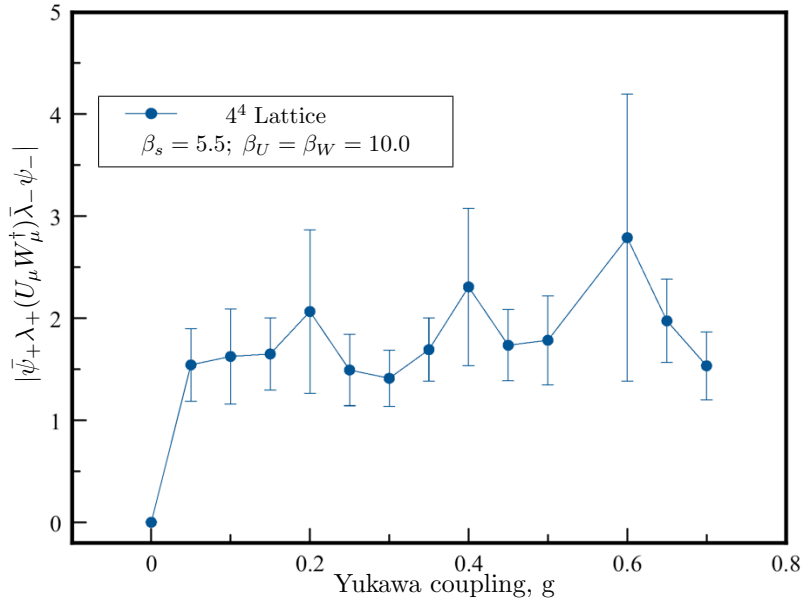}\end{center}
\caption{\label{nonsterile2} Absolute value of the four fermion condensate vs $g$ with $\beta_{S} = 5.5$ for the case
of breaking to the diagonal subgroup}
\end{figure}

\begin{figure}
\begin{center}
\includegraphics[height=95mm]{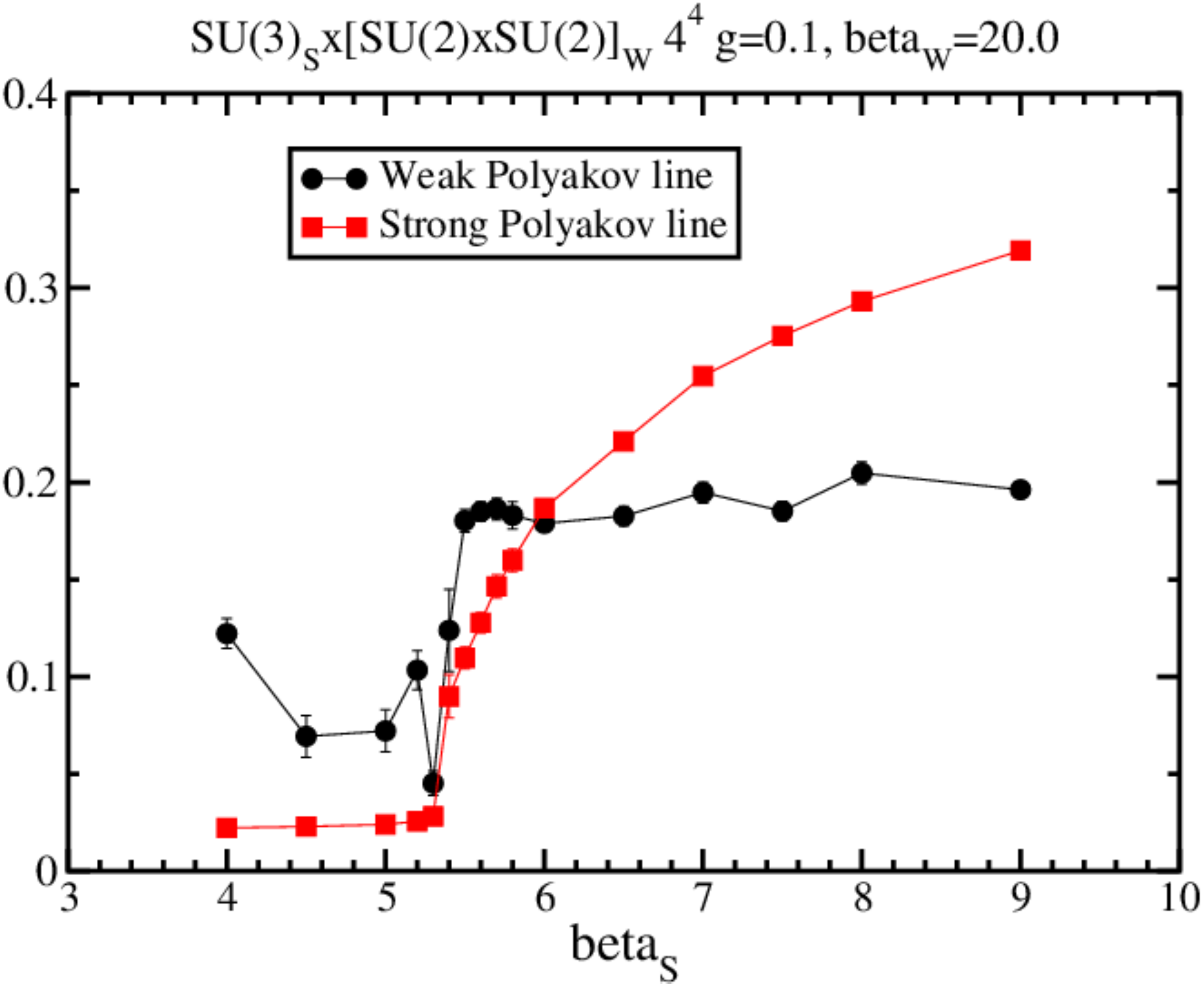}\end{center}
\caption{\label{nonsterile3} Absolute value of the strong (squares)  and weak (circles) Polyakov lines vs $\beta_S$ with $g = 0.1$}
\end{figure}

We now examine the situation when the weak gauge coupling is switched on $\beta_W=10.0$.
As a first step, we analyzed the behavior of the gauge invariant four fermion term given by
\beq
\sum_\mu
g<\psib_+(x)\lambda_+(x)>U_\mu(x)W^{*}_{\mu}(x)<\lambdab_-(x+\mu)\psi_-(x+\mu)>. \eeq
This four fermion condensate, while strictly not an order parameter, is expected to be enhanced in the Higgs phase and this is indeed what we see in fig.~\ref{nonsterile1}. 
Clearly for $\beta_S > 6.0$ the condensate is small and approaches zero while for $\beta_S< 6.0$ it grows and approaches
a plateau within errors (we fix the weak coupling $\beta_W=10.0$ throughout the numerical work).
Notice that the plot in fig.~\ref{nonsterile1} was constructed with an auxiliary coupling $g=0.1$. It is also very important to see how the condensate behaves
for fixed strong coupling $\beta_S$ as we vary $g$. Fig.~\ref{nonsterile2} shows a plot at $\beta_S=5.5$ as we scan in $g$. Clearly the value of the
condensate is insensitive to the auxiliary coupling  until $g$ becomes very small. We conclude that the magnitude of the four
fermion condensate is determined primarily by the value of the strong coupling constant as expected.
Further evidence that the lattice is indeed in a Higgs phase can be derived by plotting the appropriate weak Polyakov line that would
correspond to the free energy of a static source transforming as a bi-fundamental under the weak symmetries.
Explicitly we consider
\beq
P_W(x)=\frac{1}{4}\Tr\prod_t U_t(x,t)\Tr\prod_t W^*_t(x,t)
\eeq
 This is shown  in
fig.~\ref{nonsterile3} as a function of the strong coupling constant (again we fix the auxiliary Yukawa coupling $g=0.1$ and this
time fix $\beta_W=20.0$). The plot also shows for
reference the usual Polyakov line corresponding to the strongly coupled $SU(3)$ gauge field.
The behavior of the latter corresponds to what is expected for a system with a confining, chirally broken phase
for small $\beta_S$ and a
deconfined phase at large $\beta_S$. The weak Polyakov line is more interesting; for large $\beta_S$ it is approximately constant and relatively large consistent
with
a deconfined phase as would be expected for such a large value of the bare inverse coupling $\beta_W=10.0$ on a small volume. However, precisely
where the latter shows evidence for confinement the weak line
changes dramatically falling swiftly to smaller values.  We can interpret
this as a change in the free energy of 
an isolated fermion charged under the weak gauge groups once the latter symmetries are HIggsed by virtue of the
four fermion condensate.

\section{Discussion}

In this paper we have shown how a model based on reduced staggered fermions can be constructed that spontaneously breaks an exact global $SU(2)$ symmetry
as a result of strong dynamics. Remarkably, while the $SU(2)$ enters the theory as a vector symmetry acting on the staggered
lattice fields, we show that it should be interpreted in the continuum limit as an
axial symmetry of Dirac fermions. According to the Vafa Witten theorem this is a necessary condition for this 
broken phase to survive the
continuum limit.  Furthermore since the
continuum theory is vector-like this weak $SU(2)$ symmetry may then be gauged without
generating anomalies \cite{Nielson,KS} and  one would expect the broken symmetry phase to reappear as a Higgs phase.
In support of these theoretical arguments  we see numerical evidence of an enhancement
in the four fermion condensate and a dramatic change in the  weak Polyakov line once the strong sector confines and breaks
chiral symmetry.
We stress that the correct gauge invariant
description of the dynamical Higgs mechanism is that it proceeds via condensation of an entirely gauge
invariant four fermion operator built from strong interaction singlets\footnote{see \cite{Napoly} for a similar conclusion in a different model}..

While we have provided strong arguments and numerical evidence in favor of a Higgs phase at non-zero lattice spacing it is less clear what
happens in the continuum limit. Indeed, if one assumes universality, one should be able to replace the lattice model with a continuum
theory containing only Dirac fermions.  In such a vector-like theory it is always possible to write down fully gauge invariant fermion bilinear
terms and in principle one might have expected the vacuum of the system to break no gauge symmetries whatsoever. In the limit
in which the weak coupling is sent to zero there is no contradiction between our results and this picture; in that limit the vacuum is highly
degenerate corresponding to the many ways of pairing the elementary fermions into condensates. In some of these condensates the
weak symmetries are broken  and in some they are not but since they are all physically equivalent this is irrelevant. However, once the
weak gauge coupling is non-zero this vacuum degeneracy is broken and  question of  which strong interaction vacuum now
yields the true lowest energy state is termed the
vacuum alignment problem. We thank Maarten Golterman and Yigal Shamir for pointing this out to us and for sharing
their notes \cite{MS}. The usual folklore is that in a vector-like continuum theory the
system will adopt a vacuum in which the weak gauge interactions are left unbroken \cite{Peskin:1980gc,Banks:1984gj}. 

In our lattice model the situation is less clear;  the strong interaction vacuum state is already unique at non-zero lattice spacing and corresponds to a broken weak $SU(2)$
symmetry.  We expect for non zero lattice spacing that this vacuum is not disturbed for sufficiently weak $SU(2)$ gauge coupling and hence
the lattice theory should be found in a Higgs phase. Indeed, our numerical results are consistent with this
picture. The question of what happens in the continuum limit is then somewhat subtle; the
weak interactions presumably favor a vacuum in which $SU(2)$ is left unbroken but this will cost an energy arising from strong interactions - the
gauge invariant single link condensate costs an energy relative to the single site condensate at non-zero lattice spacing.
Presumably the correct way to
obtain a continuum limit is to send both $\beta_S\to\infty$  {\it and} $\beta_W\to\infty$ holding $\beta_S/\beta_W=r$ with $r={\rm constant}<<1$. We know
that at $r=0$ the symmetry is broken and it is logically possible that this survives to small $r$.
This limit may be explored using numerical simulation.  
At a minimum our results suggest that the symmetric vacuum favored by
universality may not be seen until simulations reach very small lattice spacings so that the lattice theory we
have described may therefore serve as a good low energy effective field theory
for studying dynamical gauge symmetry breaking.

The observation that
lattice vector symmetries can yield broken continuum axial symmetries in the presence of a non-perturbative condensate
is intriguing and deserves further study. It can be seen in a more familiar example; staggered quark formulations of QCD. The massless
staggered theory possesses a $U(1)\times U(1)$ symmetry corresponding to independent  vector $U(1)$ rotations of the two
reduced fields that make up the complete staggered field.
The usual single site condensate that forms will break this to the diagonal $U(1)$. Following the same arguments as used in section~\ref{global} one can show
that this broken $U(1)$ generator acquires a factor of $\gamma_5$ in the continuum limit where it can be interpreted as giving rise to
an axial symmetry. Indeed this broken $U(1)$ is nothing more than the usual $U(1)_A$ of staggered fermions.
In this case however the lattice fermion measure is {\it not} invariant under a {\it local} $U(1)$ transformation and so this
symmetry {\it cannot} be gauged. This is consistent with the existence of the continuum  axial anomaly.

Clearly it is important to study this model further on larger lattices so that the continuum limit can be monitored  more carefully. We hope to report on
such a study in the future. 

\section*{Acknowledgments}
The authors would like to thank Poul Damgaard for 
useful conversations and 
the DOE for partial support
under grant DE-SC0009998. The simulations were carried out using USQCD resources
at Fermilab.

\end{document}